\documentclass[3p]{elsarticle}

\usepackage{lineno,hyperref}
\modulolinenumbers[5]

\usepackage{amsmath}
\usepackage{amssymb}
\usepackage{color}
\usepackage{xcolor}
\usepackage{todonotes}
\usepackage{ulem}

\newcommand{\Neff}{\ensuremath{N_{\rm eff}}}

\begin{document}

\begin{frontmatter}


\title{Cosmological radiation density with\\ non-standard neutrino-electron interactions}

\author[Sweden]{Pablo F.\ de Salas}
\ead{pablo.fernandez@fysik.su.se}
\author[Torino]{Stefano Gariazzo}
\ead{gariazzo@to.infn.it}
\author[IFIC,FTUV]{Pablo Mart\'{\i}nez-Mirav\'e\corref{ca}}
\cortext[ca]{Corresponding author}
\ead{pamarmi@ific.uv.es}
\author[IFIC]{Sergio Pastor}
\ead{pastor@ific.uv.es}
\author[IFIC,FTUV]{Mariam T\'ortola}
\ead{mariam@ific.uv.es}

\address[Sweden]{The Oskar Klein Centre for Cosmoparticle Physics\\
Department of Physics, Stockholm University, SE-106 91 Stockholm, Sweden}
\address[Torino]{INFN, Sezione di Torino, Via P. Giuria 1, I--10125 Torino, Italy}
\address[IFIC]{Instituto de F{\'\i}sica Corpuscular  (CSIC-Universitat de Val{\`e}ncia)\\ 
	Parc Cient{\'\i}fic UV, C/ Catedr{\'a}tico Jos{\'e} Beltr{\'a}n, 2, E-46980 Paterna, Spain}
\address[FTUV]{Departament de F{\'\i}sica Te\`orica, Universitat de Val{\`e}ncia, 46100 Burjassot, Spain}

\begin{abstract}
Neutrino non-standard interactions (NSI) with electrons are known to alter the picture of neutrino decoupling from the cosmic plasma. NSI modify both flavour oscillations through matter effects, and the annihilation and scattering between neutrinos and electrons and positrons in the thermal plasma.
In view of the forthcoming cosmological observations, we perform a precision study of the impact of non-universal and flavour-changing NSI on the effective number of neutrinos, $N_{\rm eff}$.
We present the variation of $N_{\rm eff}$ arising from the different NSI parameters and discuss the existing degeneracies among them, from cosmology alone and in relation to the current bounds from terrestrial experiments. Even though cosmology is generally less sensitive to NSI than these experiments, we find that future cosmological data would provide competitive and complementary constraints for some of the couplings and their combinations.
\end{abstract}


\end{frontmatter}


\section{Introduction}

The current data from solar, atmospheric, reactor and accelerator neutrino experiments are well explained by the existence of flavour oscillations
in the framework of three-neutrino mixing (see e.g.\ \cite{deSalas:2020pgw}). This experimental evidence for non-zero neutrino masses and mixing calls for new physics beyond the Standard Model (SM) of fundamental particles and forces. Many extended theoretical models predict the existence of non-standard interactions (NSI) of neutrinos with other fermions and among themselves. If indeed neutrinos did not only interact with matter through the SM weak processes, these NSI could leave their imprint in a variety of experiments. On the one hand, they would affect the results of oscillation experiments, modifying neutrino production and detection, as well as propagation in a medium through matter effects. On the other hand, non-standard interactions would also influence other measurements, such as those from neutrino scattering experiments. Combining all available data, there is no evidence for the presence of NSI and the corresponding bounds can be derived, as summarised e.g.\ in the review \cite{Farzan:2017xzy}. Non-zero NSI could also affect astrophysical and cosmological scenarios involving neutrinos. Here we consider their implications on the decoupling process of neutrinos in the early universe, as noted in \cite{Berezhiani:2001rs,Davidson:2003ha}. 

It is well known that weak interactions keeping neutrinos 
in equilibrium with the primeval plasma become ineffective
at a temperature of ${\cal O}$(MeV)
and these elusive particles decouple, constituting the cosmological neutrino background. This process is nowadays well understood, and the percent deviations of the neutrino energy distributions from a thermal spectrum have been very precisely calculated \cite{Akita:2020szl,Froustey:2020mcq,Bennett:2020zkv}, including all relevant effects: neutrino collisions with SM interactions, finite-temperature
corrections to quantum electrodynamics (QED)
and flavour neutrino oscillations.
The corresponding value of the effective number of neutrinos is found to be $\Neff=3.044$, where $\Neff$ is a parameter related to the cosmological density of radiation ($\rho_{\rm rad}$) that quantifies the ratio of energy densities of neutrino-like relics to photons ($\rho_\gamma$) \cite{Lesgourgues:2018ncw},
\begin{equation}
\rho_{\rm rad} = \rho_\gamma \left(1+\frac{7}{8}\left(\frac{4}{11}\right)^{4/3}\,\Neff\right)\, .
\label{eq:neff}
\end{equation}
Measurements of the cosmological microwave background (CMB) anisotropies by Planck, combined with other cosmological data, constrain $\Neff=2.99^{+0.34}_{-0.33}$ at 95\% C.L.~\cite{Aghanim:2018eyx}.
Such value severely restricts the possible
existence of additional relativistic particles.

Non-standard interactions with electrons, the only charged leptons that are still abundant at MeV cosmological temperatures, would modify the thermal contact of neutrinos with the QED plasma and change their momentum spectra, leading to a different value of $\Neff$. Earlier analyses of the effect of NSI on \Neff\ include \cite{Mangano:2006ar,deSalas:2016ztq}, but these works only considered some particular choices of the NSI parameters. 

Prompted by the present allowed range of \Neff\ and the forecast sensitivities of forthcoming CMB observations in this decade, such as the Simons Observatory ($\sigma(\Neff)\simeq 0.05-0.07$) \cite{Ade:2018sbj} or the CMB-S4 project ($\sigma(\Neff)\simeq 0.02-0.03$) \cite{Abazajian:2016yjj}, in this letter we revisit the process of neutrino decoupling with NSI. We extend previous analyses of Refs.~\cite{Mangano:2006ar,deSalas:2016ztq} by computing the effect for various ranges of NSI parameters, both for one-at-a-time and multi-parameter choices. Our calculations are performed with a modified version of the
precision neutrino decoupling code {\tt FortEPiaNO} \cite{Gariazzo:2019gyi}, that now includes all relevant effects for a four-digit calculation of \Neff, as described in \cite{Bennett:2020zkv}.
We discuss how future cosmological measurements of \Neff\ can be used as a complementary way to constrain  these exotic scenarios.

\section{Neutrino non-standard interactions with electrons}

Given that the main interest is to study the decoupling of relic neutrinos, only neutral-current neutrino non-standard interactions (NC-NSI) with electrons will be considered. The effective Lagrangian including NSI and SM interaction reads

\begin{equation}
        \mathcal{L} =\mathcal{L}_{SM} + \mathcal{L}_{NSIe},
\end{equation}
where we have defined
\begin{subequations}
\begin{equation}
        \mathcal{L}_{SM}= - 2\sqrt{2}\,G_F \left[ \left(\overline{\nu}_e \gamma^\mu P_{L} e \right) (\overline{e}\gamma_\mu P_{L} \nu_e) + \sum_{X, \alpha} g_{X}\left(\overline{\nu}_\alpha \gamma^\mu P_L \nu_\alpha \right) (\overline{e}\gamma_\mu P_{X} e) \right],
\label{eq:sm}
\end{equation}

\begin{equation}
    \label{eq:nsi}
    \mathcal{L}_{NSIe} = -2\sqrt{2}\,G_F \sum_{\alpha, \beta} \varepsilon_{\alpha \beta}^{X} \left(\overline{\nu}_\alpha \gamma^\mu P_L \nu_\beta \right) (\overline{e}\gamma_\mu P_{X} e) \, .
\end{equation}
\end{subequations}

In the expressions above,  $G_F$ is the Fermi constant and the index $X = \lbrace L,R \rbrace$, so that $P_X$ denotes the chiral projectors $P_{R,L} = \left(1 \pm \gamma_{5}\right)/2$. The strength of the (neutral current) weak interactions is set by the couplings $g_L=\sin^2\theta_W-1/2$ and $g_R=\sin^2\theta_W$, with $\theta_W$ being the weak mixing angle.
Greek indices refer to the different flavours: $\alpha , \beta = e, \mu, \tau$. In the effective Lagragian for NSI (eq.~\eqref{eq:nsi}), the dimensionless coefficients $\varepsilon_{\alpha \beta}^{X}$ parametrise the strength of the interaction between flavours $\alpha$ and $\beta$.
Alternatively, this Lagrangian can be expressed in terms of the axial and vector contributions as follows:
\begin{equation}
    \mathcal{L}_{NSIe} = -2\sqrt{2}\,G_F \left[\sum_{\alpha, \beta} \varepsilon_{\alpha \beta}^{V} \left(\overline{\nu}_\alpha \gamma^\mu L \nu_\beta \right) (\overline{e}\gamma_\mu e) + \sum_{\alpha, \beta} \varepsilon_{\alpha \beta}^{A} \left(\overline{\nu}_\alpha \gamma^\mu L \nu_\beta \right) (\overline{e}\gamma_\mu \gamma_5 e) \right],
\label{eq:vansi}
\end{equation}
where we have introduced the dimensionless parameters
\begin{equation}
    \varepsilon_{\alpha \beta}^{V} \equiv \varepsilon_{\alpha \beta} ^{R} + \varepsilon_{\alpha \beta} ^{L} \, ,  \quad \text{and} \quad
    \varepsilon_{\alpha \beta}^{A} \equiv \varepsilon_{\alpha \beta} ^{R} - \varepsilon_{\alpha \beta} ^{L}.
    \label{eq:eps_VA}
\end{equation}
These definitions will become relevant when we discuss the impact of NSI on neutrino decoupling including flavour oscillations.

Non-standard interactions can be further classified from the point of view of the symmetries of the Standard Model. For non-zero values of $\varepsilon_{\alpha \beta}^{X}$, with $\alpha \neq \beta$, lepton flavour symmetry is no longer conserved and, consequently, they are often referred as \textit{flavour-changing NSI}. In the case of $\alpha = \beta$, if the difference $\varepsilon_{\alpha \alpha}^{X} - \varepsilon_{\beta \beta}^{X} \neq 0$, lepton flavour universality does not hold anymore. Hence, such interactions are the so-called \textit{non-universal NSI}.

\section{Current bounds on NSI with electrons}

Current bounds on neutrino non-standard interactions with electrons come from terrestrial experiments. They rely on the detailed analysis of neutrino  oscillation experiments and on cross-section measurements performed in scattering experiments, but they are also complemented by the study of several observables at LEP.

Neutrino non-standard interactions with electrons are known to modify the standard neutrino oscillation picture. Non-observation of the predicted deviations have allowed us to set constraints on  the NSI parameters.  Neutrino oscillations are sensitive to the vectorial part of the NSI interaction via matter effects in the neutrino propagation. Likewise,  NSI  may modify the cross sections of the detection processes employed in the experiments. In that case, neutrino experiments are sensitive to both sets of chiral NSI parameters, $\varepsilon_{\alpha \beta}^{L}$ and $\varepsilon_{\alpha\beta}^{R}$. For instance, data from solar experiments and KamLAND have been used to explore the parameter space involving non-universal and/or flavour-changing NSI \cite{Miranda:2004nb,Bolanos:2008km,Escrihuela:2010zz,Agarwalla:2012wf, Khan:2017oxw}. Alternatively, long-baseline and atmospheric experiments, which are also sensitive to NSI, were used to constrain the vector parameters $|\varepsilon_{\tau \tau}^{V} - \varepsilon_{\mu \mu }^{V}|$ and $|\varepsilon_{\mu \tau}^{V}|$ \citep{GonzalezGarcia:2011my, Salvado:2016uqu, Demidov:2019okm}. 

The accurate measurement of the cross-section of purely leptonic processes like neutrino-electron scattering also allows us to constrain the NSI parameters. The TEXONO Collaboration studied electron antineutrino scattering on electrons ($\overline{\nu}_e e \rightarrow \overline{\nu}_e e$) \cite{Deniz:2010mp} like other reactor experiments such as Irvine \cite{PhysRevLett.37.315}, MUNU \cite{Daraktchieva:2003dr} and Rovno \cite{Derbin:1993wy} had previously done. The LSND experiment performed measurements of electron neutrino scattering on electrons ($\nu_e e \rightarrow \nu_e e$) \cite{Davidson:2003ha,PhysRevD.63.112001}. The combination of both experiments can set constraints on the NSI parameters involved: $\varepsilon^L_{ee}$, $\varepsilon^R_{ee}$, $\varepsilon^L_{e\mu}$, $\varepsilon^R_{e\mu}$, $\varepsilon^L_{e\tau}$ and $\varepsilon^R_{e\tau}$ \cite{Barranco:2005ps, Khan:2016uon}. In addition, precise measurements of muon neutrino scattering on electrons ($\overline{\nu}_\mu e \rightarrow \overline{\nu}_\mu e$ and $\nu_\mu e \rightarrow \nu_\mu e$) by the CHARM Collaboration set very stringent constraints on $\varepsilon^L_{\mu \mu}$, $\varepsilon^R_{\mu \mu}$, $\varepsilon^L_{e\mu}$, $\varepsilon^R_{e\mu}$, $\varepsilon^L_{\mu\tau}$ and $\varepsilon^R_{\mu\tau}$ \cite{Vilain:1994qy}.

A degeneracy present in the bounds on non-universal NSI from scattering experiments can be resolved with the inclusion of the measurement of the forward-backward asymmetry in $e^{+} e^{- }\rightarrow e^+ e^- $ at LEP. On top of that, one can also study the reaction $e^{+} e^{- }\rightarrow \nu \overline{\nu} \gamma $, mediated by the $Z$ and $W$ bosons in the SM.
Further constraints on neutrino-electron NSI can be placed from additional coherent contributions in the presence of NSI, which modify the expected number of events \cite{Barranco:2007ej}. This process has been studied by the four collaborations operating at LEP.

The current bounds on neutrino non-standard interactions with electrons, derived considering only one non-zero parameter at a time, are summarised in Table \ref{tab:nu} and Table \ref{tab:fc} for non-universal and flavour-changing NSI, respectively.

\begin{table}[]
\centering
\begin{tabular}{|c|c|}
\hline
Parameter and $90\%$ C.L range & Origin \\ \hline
  -0.021  $ <  \varepsilon^L_{ee} < $  0.052 & Neutrino oscillations \citep{Bolanos:2008km} \\ \hline
-0.07  $< \varepsilon^R_{ee} <$  0.08   & Neutrino scattering \cite{Deniz:2010mp} \\ \hline
-0.03 $ < \varepsilon^L_{\mu\mu}$, $\varepsilon^R_{\mu\mu} < $ 0.03 &  Neutrino scattering and accelerator data \cite{Barranco:2007ej}\\ \hline
-0.12  $< \varepsilon^L_{\tau\tau} <$ 0.06  & Neutrino oscillations \citep{Bolanos:2008km} \\ \hline
-0.98  $< \varepsilon^R_{\tau\tau} <$  0.23 &  Neutrino oscillation \cite{Bolanos:2008km, Agarwalla:2012wf} \\
-0.25  $< \varepsilon^R_{\tau\tau} <$  0.43  & Neutrino scattering and accelerator data \citep{Bolanos:2008km}\\ \hline
\end{tabular}
\caption{Current bounds on non-universal NSI at 90$\%$ C.L. for 1 degree of freedom. Adapted from Ref.\ \cite{Farzan:2017xzy}. 
}
\label{tab:nu}
\end{table}

\begin{table}[]
\centering
\begin{tabular}{|c|c|}
\hline
Parameter and $90\%$ C.L range & Origin \\ \hline
  -0.13  $ <  \varepsilon^L_{e\mu}, \varepsilon^R_{e\mu} < $  0.13 & Neutrino scattering and accelerator data \cite{Barranco:2007ej}\\ \hline
 -0.33  $< \varepsilon^L_{e\tau} <$  0.33   & Neutrino scattering and accelerator data \cite{Barranco:2007ej} \\ \hline
-0.28 $ < \varepsilon^R_{e\tau} < $ -0.05 $\&$ 0.05 $ < \varepsilon^R_{e\tau} < $ 0.28 &  Neutrino scattering and accelerator data \cite{Barranco:2007ej}\\ 
-0.19 $ < \varepsilon^R_{e\tau} < $ 0.19 & Neutrino scattering \cite{Deniz:2010mp} \\ \hline
-0.10  $< \varepsilon^L_{\mu\tau}, \varepsilon^R_{\mu\tau} <$ 0.10  & Neutrino scattering and accelerator data \cite{Barranco:2007ej}\\  \hline
\end{tabular}
\caption{Current bounds on flavour-changing NSI at 90$\%$ C.L. for 1 degree of freedom. Adapted from Ref.\ \cite{Farzan:2017xzy}.}
\label{tab:fc}
\end{table}

\section{Impact of non-standard interactions on \Neff}

Non-standard interactions with electrons also affect the calculation of the effective number of neutrinos,
which would deviate from the standard value $\Neff=3.044$ \cite{Froustey:2020mcq,Bennett:2020zkv}.
The effect of NSI enters the calculation of the neutrino thermalisation in the early universe in two different ways: (i)
through the matter effects that alter neutrino oscillations in the dense electromagnetic plasma, and (ii) through the collision integrals that encode the scattering and annihilation between neutrinos and electrons.
In \cite{Bennett:2020zkv}, the authors showed that neutrino oscillations change \Neff\ in less than 0.001, much below the expected experimental sensitivity of the incoming generation of cosmological observations.
Consequently, we expect the impact of NSI through matter effects to be much smaller than the modification they cause on the interactions of neutrinos with electrons.

In our calculation, we use the density matrix formalism, where the diagonal elements of the neutrino density matrix $\varrho$ are the momentum distribution functions of the different neutrino flavours, and the off-diagonal elements implement the coherence of the system. We use the comoving variables adopted in previous papers (see e.g.\ \cite{Bennett:2020zkv,Gariazzo:2019gyi}): $x=m_e a$, $y=p a$, $z=T_\gamma a$; where $m_e$ is the electron mass, $p$ the neutrino momentum, $T_\gamma$ the photon temperature and $a$ the cosmological scale factor. Written in terms of these variables, the evolution of $\varrho$ can be computed according to the following equation:
\begin{equation}
\label{eq:drho_dx_nxn}
\frac{{\rm d}\varrho(y)}{{\rm d}x}
=
\sqrt{\frac{3 m^2_{\rm Pl}}{8\pi\rho}}
\left\{
    -i \frac{x^2}{m_e^3}
    \left[
        \frac{\mathbb{M}_{\rm F}}{2y}
        -
        \frac{2\sqrt{2}G_{\rm F} y m_e^6}{x^6}
        \left(
            \frac{\mathbb{E}_\ell+\mathbb{P}_\ell}{m_W^2}
            +
            \frac{4}{3}\,\frac{\mathbb{E}_\nu}{m_Z^2}
        \right),
    \varrho
    \right]
    +\frac{m_e^3}{x^4}\mathcal{I(\varrho)}
\right\}\,,
\end{equation}
where
$m_{\rm Pl}$ is the Planck mass,
$m_W$ and $m_Z$ the masses of the $W$ and $Z$ bosons,
$\rho$ is the total energy density of the universe,
$\mathbb{M}_{\rm F}$ is the rotated neutrino mass matrix,
$\mathbb{E}_\ell$, $\mathbb{P}_\ell$ and $\mathbb{E}_\nu$ represent the energy density and pressure of charged leptons (see below) and neutrinos,
and $\mathcal{I(\varrho)}$ encodes the collision terms.
These are the sum of two contributions, one including neutrino-electron scattering and annihilation, and the other related to neutrino-neutrino interactions, which are thoroughly treated in \cite{Bennett:2020zkv} and will not be discussed here in full details.

Neutrino-electron interactions can be computed as a multidimensional integral,
which can be reduced to a two-dimensional integral (see e.g.~\cite{deSalas:2016ztq}).
We do not explicitly write the full expression for these integrals here, the interested reader can find them in \cite{Bennett:2020zkv}.
For the sake of understanding the impact of NSI on \Neff,
we only need to remember that the collision integrals are functions of the following structure
\begin{eqnarray}
\mathcal{I}_{\rm ann}
&\propto&
G_F ^2\int {\rm d}y_2 {\rm d}y_3
\sum_{a=L,R}\sum_{b=L,R}
A_{ab}(x,y,y_2,y_3)
F_{\rm ann}^{ab}\left(\varrho^{(1)}, \varrho^{(2)}, f_e^{(3)}, f_e^{(4)}\right)
\,,
\\
\mathcal{I}_{\rm sc}
&\propto&
G_F^2\int {\rm d}y_2 {\rm d}y_3
\sum_{a=L,R}\sum_{b=L,R}
B_{ab}(x,y,y_2,y_3)
F_{\rm sc}^{ab}\left(\varrho^{(1)}, f_e^{(2)}, \varrho^{(3)}, f_e^{(4)}\right)
\,,
\end{eqnarray}
where 
$a$ and $b$ represent the chirality $L$ or $R$,
$f_e$ is the electron momentum distribution function
and
the coefficients $A_{ab}$ and $B_{ab}$ do not depend on the NSI parameters.
These, instead, only enter the phase-space factors $F_{\rm ann}^{ab}$ and $F_{\rm sc}^{ab}$,
whose full expressions are reported here:
\begin{eqnarray}
F_{\rm ann}^{ab}\left(\varrho^{(1)}, \varrho^{(2)}, f_e^{(3)}, f_e^{(4)}\right)
&=&
f_e^{(3)}f_e^{(4)}\left[G^a(1-\varrho^{(2)})G^b(1-\varrho^{(1)})+(1-\varrho^{(1)})G^b(1-\varrho^{(2)})G^a\right]
\nonumber\\
&-&
(1-f_e^{(3)})(1-f_e^{(4)})\left[G^a\varrho^{(2)}G^b\varrho^{(1)}+\varrho^{(1)}G^b\varrho^{(2)}G^a\right],
\label{eq:F_ab_ann}\\
F_{\rm sc}^{ab}\left(\varrho^{(1)}, f_e^{(2)}, \varrho^{(3)}, f_e^{(4)}\right)
&=&
f_e^{(4)}(1-f_e^{(2)})\left[G^a\varrho^{(3)}G^b(1-\varrho^{(1)})+(1-\varrho^{(1)})G^b\varrho^{(3)}G^a\right]
\nonumber\\
&-&
f_e^{(2)}(1-f_e^{(4)})\left[\varrho^{(1)}G^b(1-\varrho^{(3)})G^a+G^a(1-\varrho^{(3)})G^b\varrho^{(1)}\right].
\label{eq:F_ab_sc}
\end{eqnarray}

The interactions described by the above-mentioned $G^X$ matrices 
are modified to take into account the possible additional couplings.
In presence of NSI, they read:
\begin{equation}
\label{eq:gLR_nsi}
G^L
=
\left(
\begin{array}{ccc}
\tilde g_L+\varepsilon^L_{ee} & \varepsilon^L_{e\mu} & \varepsilon^L_{e\tau}\\
\varepsilon^L_{e\mu} &  g_L + \varepsilon^L_{\mu\mu} & \varepsilon^L_{\mu\tau}\\
\varepsilon^L_{e\tau} & \varepsilon^L_{\mu\tau} & g_L + \varepsilon^L_{\tau\tau}\\
\end{array}
\right)
\qquad
\text{and}
\qquad
G^R
=
\left(
\begin{array}{ccc}
g_R+\varepsilon^R_{ee} & \varepsilon^R_{e\mu} & \varepsilon^R_{e\tau} \\
\varepsilon^R_{e\mu} & g_R + \varepsilon^R_{\mu\mu} & \varepsilon^R_{\mu\tau} \\
\varepsilon^R_{e\tau} & \varepsilon^R_{\mu\tau} & g_R + \varepsilon^R_{\tau\tau} \\
\end{array}
\right)
\,,
\end{equation}
where the diagonal elements include standard weak interactions through the couplings
$g_L$ and $g_R$ and we have adopted the shorthand notation $\tilde g_L=1+ g_L$.
The presence of NSI, therefore, modifies the phase-space factors of collision integrals.
Since the functions $F$ contain products of two $G^X$ matrices, 
interactions between neutrinos and electrons are proportional to the Standard Model coefficients $g_L^2$, $g_R^2$ and a mixed term $g_Lg_R$ (or $\tilde g_L$).
In the presence of NSI, the values of these coefficients are shifted according to
\begin{eqnarray}
g_L^{2}
&\longrightarrow&
\left(g_{L} + \varepsilon^{L}_{\alpha\alpha}\right)^2 + \sum_{\beta \neq \alpha} |\varepsilon^{L}_{\alpha \beta}|^2
\,,
\label{eq:gLsqshift}
\\
g_R^{2}
&\longrightarrow& \left(g_{R} + \varepsilon^{R}_{\alpha\alpha}\right)^2 + \sum_{\beta \neq \alpha} |\varepsilon^{R}_{\alpha \beta}|^2
\,,
\label{eq:gRsqshift}
\\
g_Lg_R
&\longrightarrow&
\left(g_L + \varepsilon^L_{\alpha\alpha} \right)\left(g_R + \varepsilon^R_{\alpha\alpha}\right) + \sum_{\beta \neq \alpha} |\varepsilon^L_{\alpha\beta}||\varepsilon^R_{ \alpha\beta}|
\,.
\label{eq:gLgRshift}
\end{eqnarray}

This means that a minimum value of \Neff\ is reached when $\varepsilon^X_{\alpha \beta} = 0$ and $\varepsilon^X_{\alpha\alpha} = - g_X$, since that corresponds to the least energy transfer through neutrino-electron interactions. 
As we mentioned above, NSI also modify  the matter effects that alter neutrino oscillations in the thermal plasma.
The modification must be included in the $\mathbb{E}_\ell$ and $\mathbb{P}_\ell$ terms that appear in equation~\eqref{eq:drho_dx_nxn}.
Defining
\begin{equation}
\label{eq:E_nsi_emu}
\mathbb{E}_e^{\rm NSI}
=
\left(
\begin{array}{cccc}
1+\varepsilon_{ee}^{V}& \varepsilon_{e\mu}^{V} & \varepsilon_{e\tau}^{V} \\
\varepsilon_{e\mu}^{V} & \varepsilon_{\mu\mu}^{V} & \varepsilon_{e\mu}^{V} \\
\varepsilon_{e\tau}^{V} & \varepsilon_{\mu\tau}^{V} & \varepsilon_{\tau\tau}^{V}
\end{array}
\right)
\,,
\end{equation}
the matrices that account for the charged lepton contribution to the matter potential of flavour oscillations can be written as
\begin{equation}
\label{eq:leptonmatterpotentials_nsi}
\mathbb{E}_\ell
=
\rho_e\,
\mathbb{E}_e^{\rm NSI}
\,,
\qquad
\qquad
\mathbb{P}_\ell
=
P_e\,
\mathbb{E}_e^{\rm NSI}
\,,
\end{equation}
where $\rho_e$ and $P_e$ are the electron energy density and pressure.
We can see that both equations include the standard term (from the 1 in the top left entry of $\mathbb{E}_e^{\rm NSI}$), plus corrections that depend on both the $L$ and $R$ NSI coefficients.
Notice that the effect of NSI in the matter potentials can be null if all the $\varepsilon^{V}_{\alpha\beta}$ vanish.

All the results we discuss in the following are obtained using the Fortran code
{\tt FortEPiaNO} (Fortran-Evolved Primordial Neutrino Oscillations\footnote{\url{https://bitbucket.org/ahep_cosmo/fortepiano_public}}) \cite{Bennett:2020zkv,Gariazzo:2019gyi}.
 We adopt numerical settings that allow an estimated precision at the level of $\lesssim5\times10^{-4}$ on the final \Neff,
more than sufficient to study the effect of NSI.
This number has been estimated by repeating the same calculation using slightly different numerical settings for the same physical parameters, and studying the behaviour of the computed \Neff.
Specifically, for all the runs that we have performed we set a numerical (absolute and relative) precision for the differential equation solver of
$10^{-6}$ or less, and an initial temperature given by $x_{\rm in}= 0.01$, a Gauss-Laguerre spacing for the neutrino momenta with $N_y= 30$ and $y_{\rm max}=20$.
Concerning collision integrals, we use the full calculations (no damping terms) for both neutrino--electron and neutrino--neutrino interactions.

In the calculation, we ignore the contribution of muons that are present in the very early universe for two reasons.
First, here we only consider NSI between neutrinos and electrons, which do not alter the interaction with muons.
Second, and most importantly, the muon density decreases exponentially much before neutrino oscillations start to be effective and neutrinos deviate from thermal equilibrium.
Muons may have a small effect on \Neff\ only when considering the presence of light sterile neutrinos, for which oscillations and the consequent thermalisation process start much before \cite{Gariazzo:2019gyi}.
We performed several numerical calculations to check whether these considerations are justified. The results showed that not including muons introduces a variation on \Neff\  slightly larger than $10^{-4}$, but still below the numerical precision aimed here.

\begin{table}[t]
    \centering
    \begin{tabular}{|c|c|c|c|c|c|c|c|}
    \hline
         $\varepsilon_{ee}^L$ & $\varepsilon_{\tau \tau}^L$ & $N_{\rm eff}$ & $N_{ \rm eff}$ - $N_{ \rm eff}^{\text{no NSI}}$ &$N_{ \rm eff}^{\text{osc}}$ & $N_{ \rm eff}^{\text{osc}}$ - $N_{ \rm eff}^{\text{no NSI}}$ & $N_{ \rm eff}^{\text{coll}}$ & $N_{ \rm eff}^{\text{coll}}$ - $N_{ \rm eff}^{\text{no NSI}}$  \\ \hline
        0.2 & -0.3 & 3.05714 & 1.4 $\times 10 ^{-2}$ & 3.04357 & -7 $\times 10 ^{-5}$ & 3.05714 &  1.4 $\times 10 ^{-2}$\\ \hline
         -0.3 & 0.2 & 3.03199 & -1.2 $\times 10 ^{-2}$ & 3.04367 & 3$\times 10^{-5}$ &3.03198 & -1.2 $\times 10 ^{-2}$ \\ \hline
    \end{tabular}
    \caption{
    \label{tab:nsi_contributions}
    Comparison between the value of $N_{\rm eff}$ obtained for two sets of NSI parameters considering only its impact on oscillations through equation~\eqref{eq:leptonmatterpotentials_nsi} ($N_{ \rm eff}^{\text{osc}}$) or in the collisional integrals through the $G^X$ matrices in equation~\eqref{eq:gLR_nsi} ($N_{ \rm eff}^{\text{coll}}$). The deviation from the value of \Neff\ expected in the absence of NSI under the same assumptions is presented as a reference, where $N_{\rm eff}^{\text{no NSI}} = 3.04364$ (muons are not included).
    }
\end{table}

As we discussed previously, NSI parameters enter the calculation of \Neff\ in two ways: they affect the neutrino-electron collision terms and they alter neutrino oscillations in the thermal plasma in the presence of a dense medium composed by electrons.
Table~\ref{tab:nsi_contributions}  shows the total and the separate effect of including NSI in the two different terms.
As we can see, the effect provided by altering neutrino oscillations is very small, as expected from the results of \cite{Bennett:2020zkv},
where the authors showed that the effect on \Neff\ from neutrino oscillations is at most of $\sim10^{-4}$ within the allowed range of mixing parameters \cite{deSalas:2020pgw}.
Remembering that for the problem at hand, we estimate fluctuations at the level of $10^{-4}$ to be compatible with numerical instability,
we could say that the effect of NSI on the oscillation term is completely negligible.
Therefore, the impact that NSI have on \Neff\ comes almost entirely from its effect on the collision terms, from which it is possible to obtain variations at the level of $10^{-2}$, to be discussed in detail in what follows.
Notice, however, that the sign of the effect provided by NSI through the collision terms is opposite with respect to their effect on neutrino oscillations.

\begin{figure}[t]
    \centering
    \includegraphics[width = 0.49\textwidth]{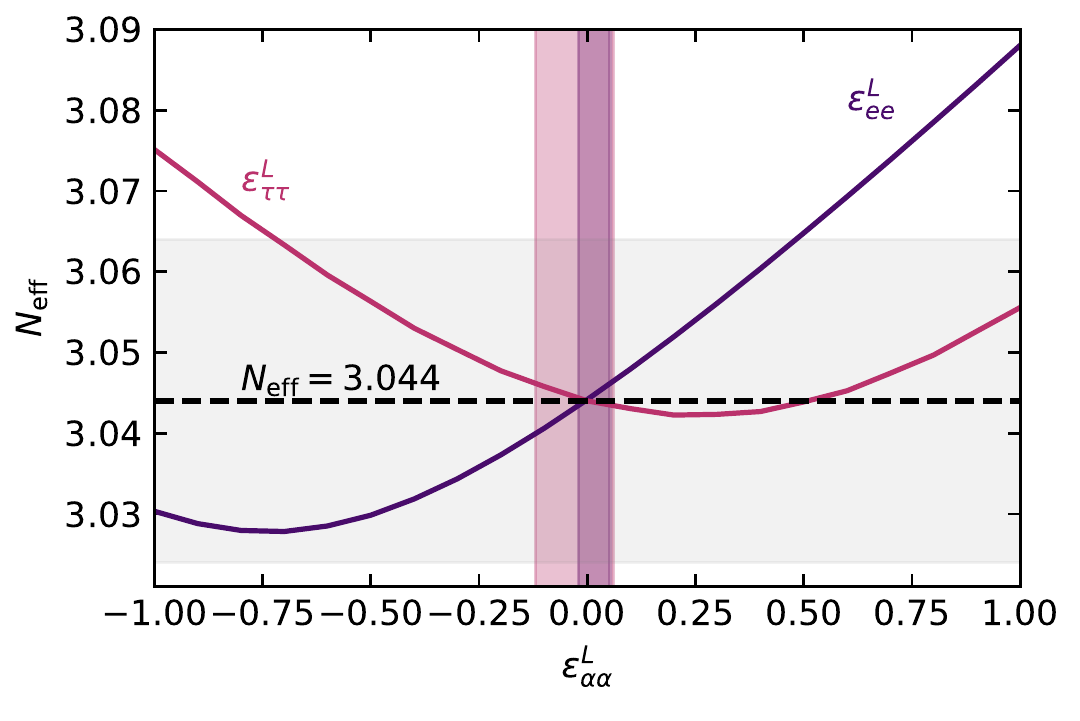}
    \includegraphics[width = 0.49\textwidth]{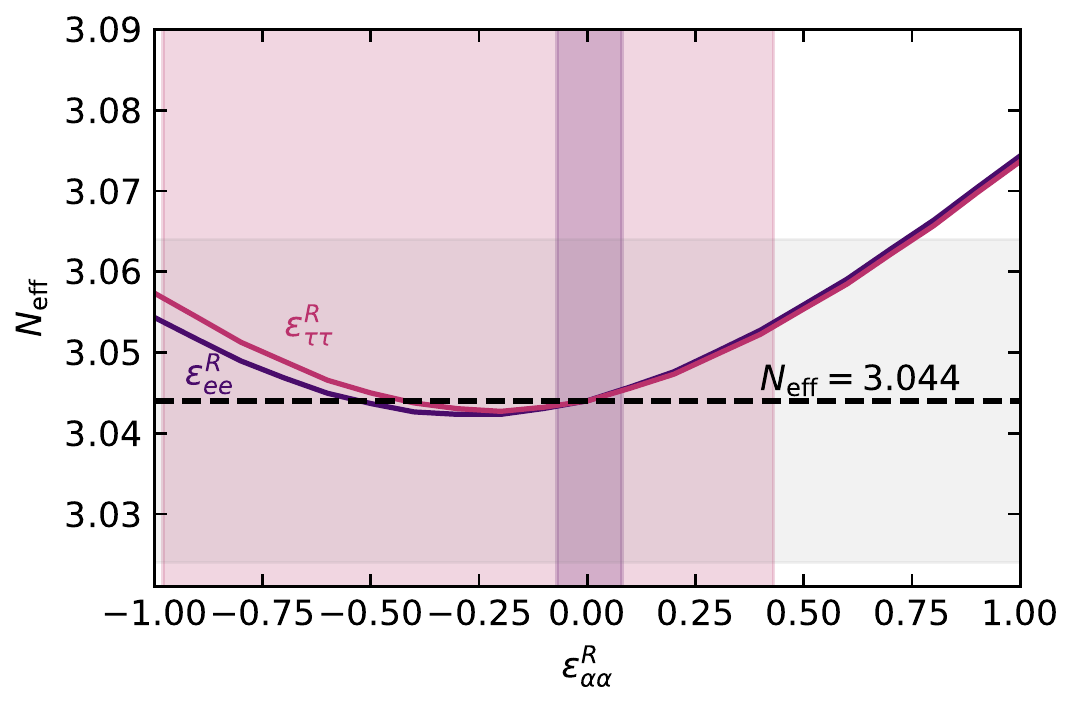}
    \caption{
    \label{fig:aaoneatatime}
    Values of $N_{\rm eff}$ in the presence of non-zero $\varepsilon^L_{\alpha \alpha}$ (left panel) and $\varepsilon^R_{\alpha \alpha}$ (right panel), for $\alpha = \lbrace e , \tau \rbrace$. The shaded vertical regions correspond to the 90$\% $ C.L. bounds presented in Table \ref{tab:nu}. The value of $N_{\rm eff} = 3.044$ expected in the absence of NSI is indicated by a dashed line, together with a shaded region corresponding to $\pm 0.02$, which is a reference value for the expected $1 \sigma$ uncertainty from future cosmological observations.
    }
\end{figure}

Concerning non-universal NSI, we present in Figure \ref{fig:aaoneatatime} the values of $\Neff$ predicted in the presence of non-zero $\varepsilon^L_{\alpha\alpha}$ or $\varepsilon^R_{\alpha\alpha}$, with $\alpha=e,\tau$.
In the left panel, one can see that for negative values of $\varepsilon^L_{ee}$ (purple line) the value of $\Neff$ decreases with respect to the Standard Model prediction, since the strength of the coupling is reduced.
The minimum, expected for
$\varepsilon^L_{ee} = -\tilde g_L = -\frac{1}{2} - \sin^2\theta_{W}$,
is visible on the left of the plot.
For the $\varepsilon^L_{\tau\tau}$ parameter (pink line), the minimum is located at $\varepsilon^L_{\tau\tau} = - g_L = \frac{1}{2} - \sin^2 \theta_W$, and it varies more slowly across the displayed range.
The trend observed in the case of the $R$ couplings (right panel) is different: for both $\varepsilon^R_{ee}$ (purple line) and $\varepsilon^R_{\tau\tau}$ (pink line), the minimum value of \Neff\ is clearly more similar for the two parameters, and it is close to $\varepsilon^R_{\alpha\alpha} = - g_R = -\sin ^2 \theta_{W}$.
The small difference between the effect of the two parameters $\varepsilon^R_{ee}$ and $\varepsilon^R_{\tau\tau}$ at large negative values arises from the mixed terms, proportional to $g_L g_R$ in the SM, since the left coupling of neutrinos to electrons or taus is different.
As a conclusion, we find that minima are found where they were expected from the argument of the shift in the coupling.
This reinforces the fact that the exact value of the oscillation parameters is not playing a major role.

From the figure, one can see that only large negative values of $\varepsilon^L_{ee}$ would lead to an effective number of neutrinos considerably smaller than $3.044$.
In spite of being experimentally constrained, it is important to note that this is one of the few scenarios that can lead to a value of $\Neff$ smaller than the Standard Model prediction.
\begin{figure}[t]
    \centering
    \includegraphics[width = 0.49\textwidth]{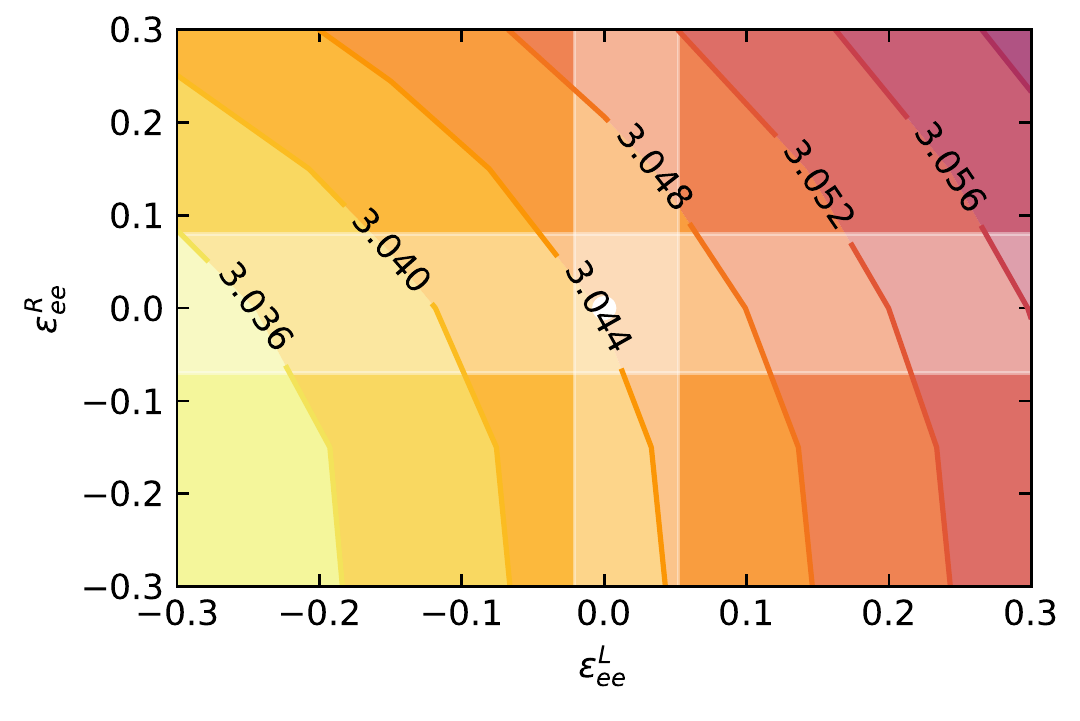}
    \includegraphics[width = 0.49\textwidth]{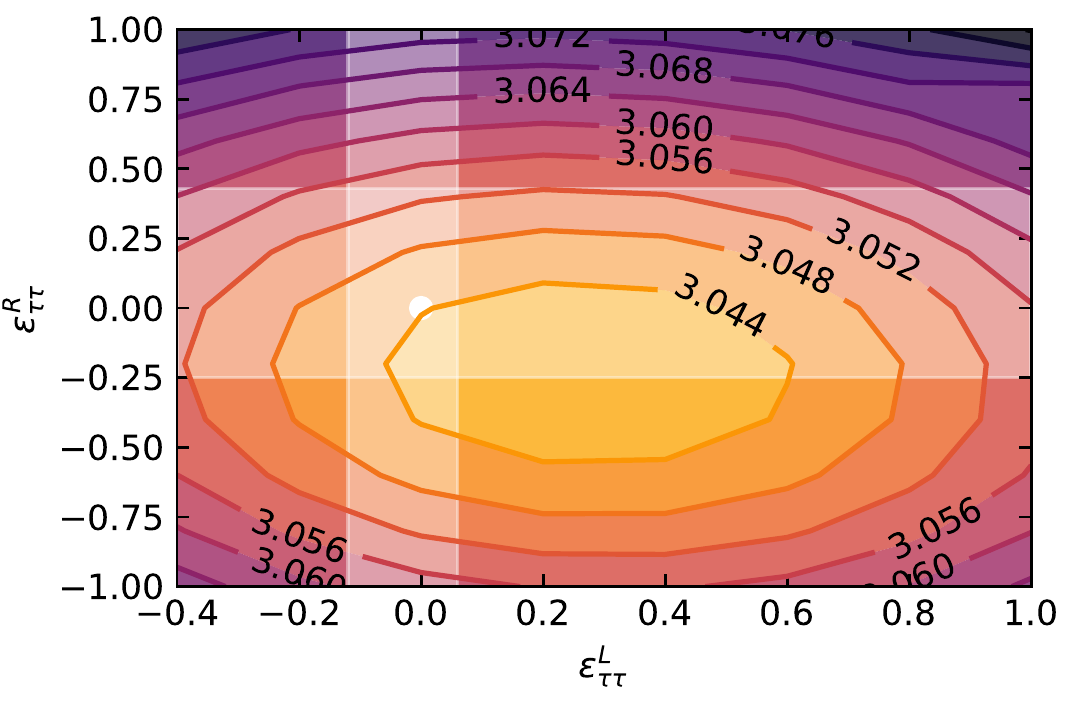}
    \includegraphics[width = 0.49\textwidth]{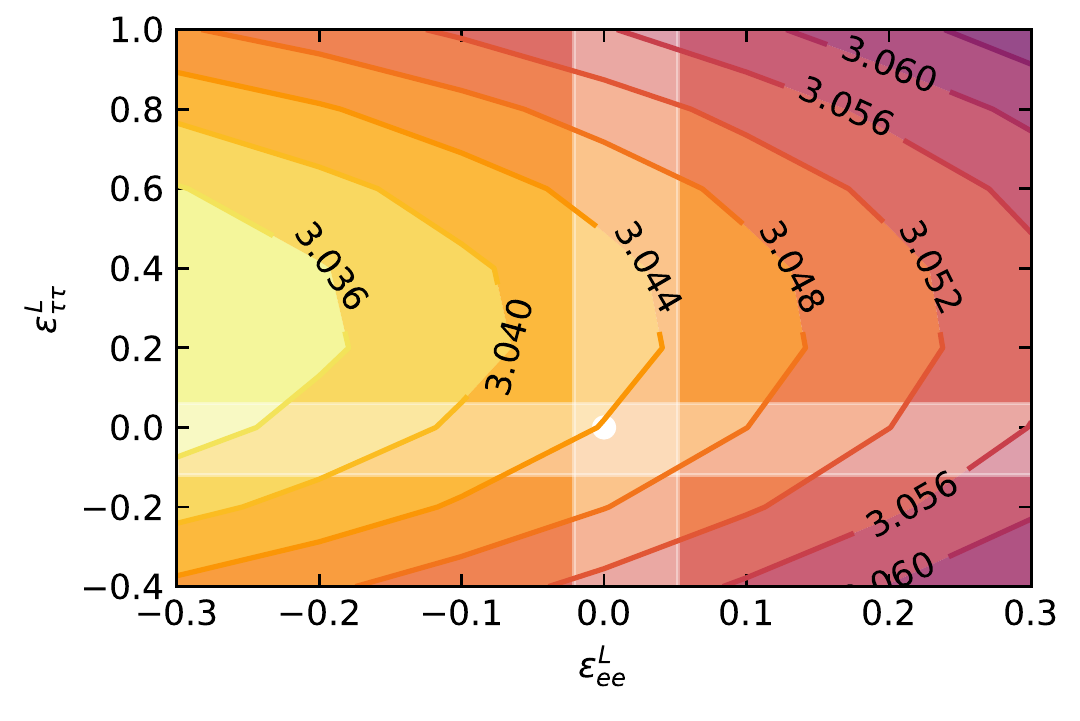}
     \includegraphics[width = 0.49\textwidth]{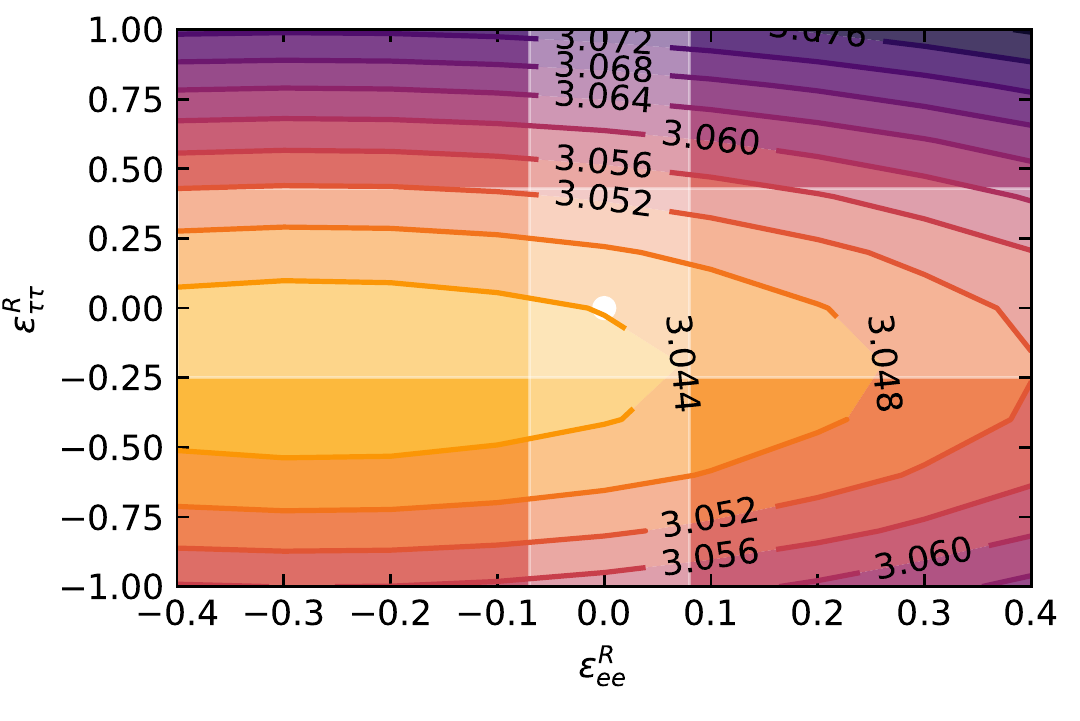}
    \caption{Values of \Neff\ when two NSI parameters are varied simultaneously: $\varepsilon^L_{ee}$- $\varepsilon^R_{ee}$ (top left panel), $\varepsilon^L_{\tau \tau}$- $\varepsilon^R_{\tau \tau}$ (top right panel), $\varepsilon^L_{ee}$- $\varepsilon^L_{\tau \tau}$  (bottom left panel) and  $\varepsilon^R_{ee}$- $\varepsilon^R_{\tau \tau}$ (bottom right panel). White-shaded regions correspond to the 90\% C.L. experimental bounds obtained varying one parameter at a time, extracted from \citep{Farzan:2017xzy}.}
    \label{fig:2dnu}
\end{figure}

It is also interesting to see the interplay between the NSI parameters once two of them are allowed to vary simultaneously.
 Figure~\ref{fig:2dnu} shows the variation of $\Neff$ induced by the simultaneous presence of two non-zero NSI couplings.
Notice that the iso-$\Neff$ contours presented in the plots are ellipses, as one can predict from the shift in the couplings entering the collisions, see equations~\eqref{eq:gLsqshift} to \eqref{eq:gLgRshift}.
Again, this confirms the fact that the impact of NSI on collisions dominates over its effect on oscillations. 
To illustrate it, one can consider that, along the line for which $\varepsilon^L_{\alpha \alpha} = - \varepsilon^R_{\alpha \alpha}$, the vectorial component of the NSI cancels out, leaving oscillations unchanged.
The change in the value of $\Neff$, therefore, arises exclusively from the shift in the value of the axial coupling entering the collision term. 

From Figures~\ref{fig:aaoneatatime} and \ref{fig:2dnu} we can also see that next-generation cosmological observations,
which are expected to determine $\Neff$ with an error that could be as small as $\pm 0.02$ \cite{Abazajian:2016yjj},
will be able to constrain non-universal NSI to the same order of magnitude of current laboratory experiments, in particular for the $\varepsilon^R_{\tau\tau}$ parameter.
Remember, however, that constraints from cosmology are indirect and, therefore, much more model-dependent than laboratory results.

\begin{figure}[t]
    \centering
    \includegraphics[width = 0.49\textwidth]{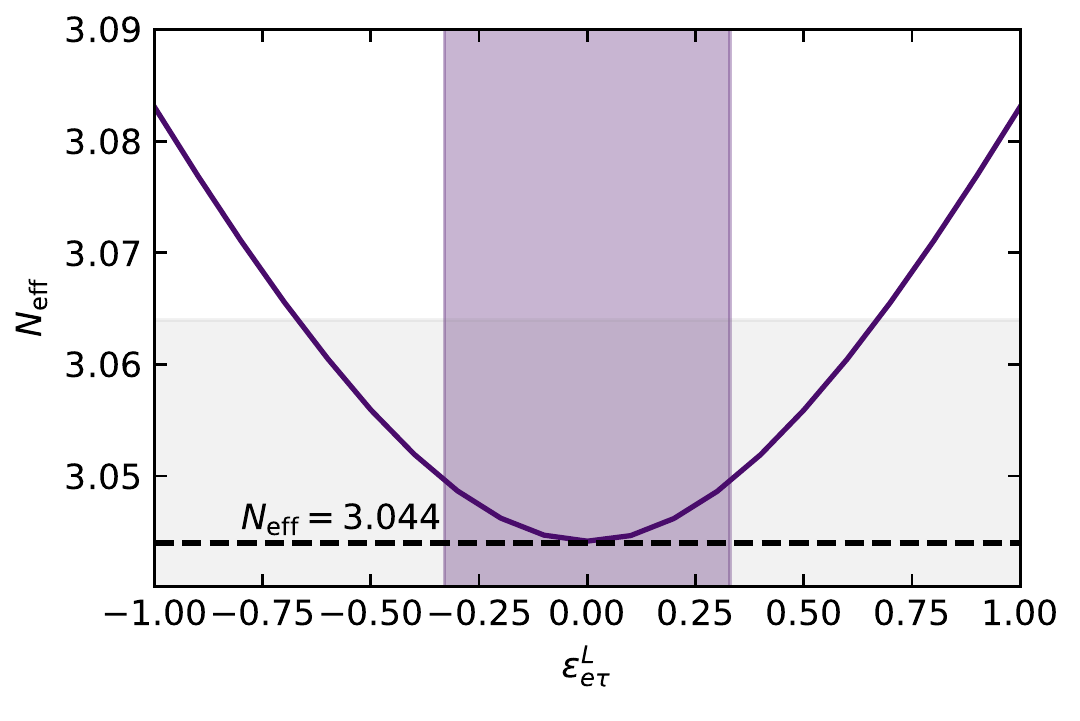}
    \includegraphics[width = 0.49\textwidth]{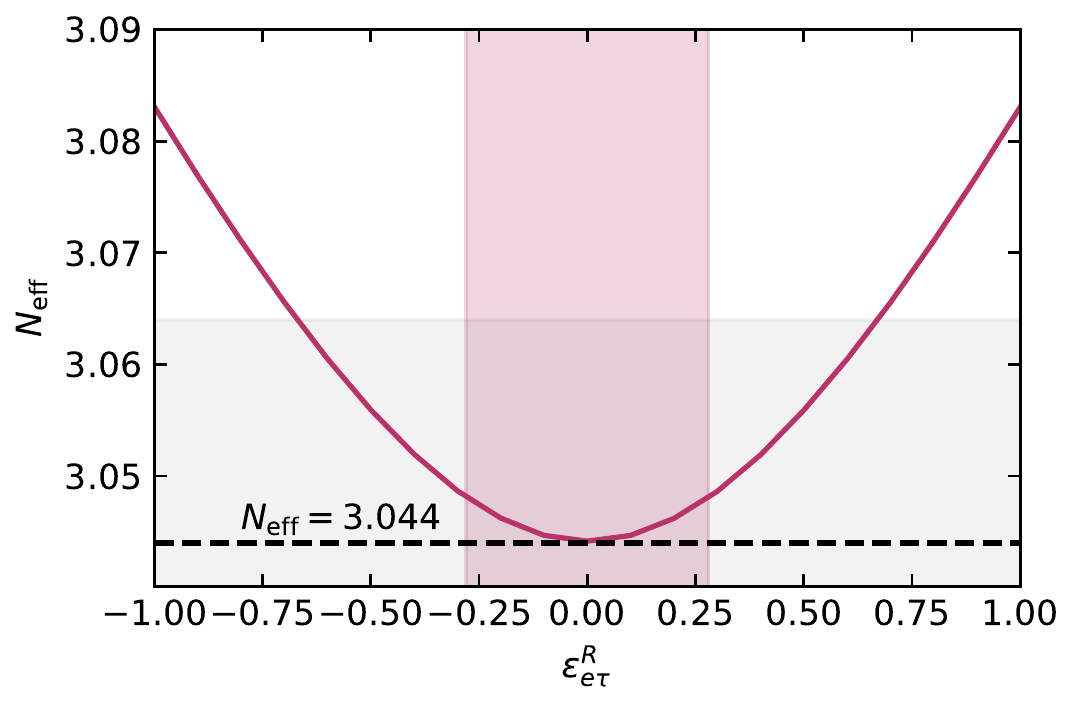}
    \caption{
    \label{fig:etoneatatime}
    Same as Fig.\ \ref{fig:aaoneatatime} for the flavour-changing NSI parameters $\varepsilon^L_{e\tau}$ (left panel) and $\varepsilon^R_{e\tau}$ (right panel).
    }
\end{figure}

Non-standard interactions leading to flavour-changing processes,
parameterised by the $\varepsilon^X_{\alpha\beta}$ ($\alpha\neq\beta$) coefficients,
can lead to higher values of $\Neff$.
This is due to the fact that they increase the strength of the interactions involving neutrinos and electrons, as we can see from equations~\eqref{eq:gLsqshift} to \eqref{eq:gLgRshift}.
Also, the effect on the collisions is independent of the sign of $\varepsilon^X_{\alpha\beta}$ and, hence, one expects the contours to be symmetric with respect to $\varepsilon^X_{\alpha\beta} = 0$.
All these considerations can be verified by looking at Figure \ref{fig:etoneatatime}, where the dependence of $\Neff$ as a function of $\varepsilon^L_{e\tau}$ (left panel) and $\varepsilon^R_{e\tau}$ (right panel) is shown.
As it happened for the non-universal NSI parameters, we can notice that future CMB constraints (horizontal gray band) would be able to constrain the flavour-changing NSI parameters to a similar level than terrestrial experiments, particularly for the $R$ component.
We repeated the same exercise also considering the $\varepsilon^L_{e\mu}$, $\varepsilon^L_{\mu\tau}$ and $\varepsilon^R_{e\mu}$, $\varepsilon^R_{\mu\tau}$ parameters.
The results are not shown in the figure because the dependence of \Neff\ on these NSI couplings is practically indistinguishable from the effect of the corresponding $\varepsilon^L_{e\tau}$ or $\varepsilon^R_{e\tau}$ parameters.
This is expected, since the effect of the different flavour-changing NSI parameters is significantly different from one another
only if the neutrino density matrix is significantly different from the identity matrix multiplied by the neutrino momentum distribution.
Given the fact that the three neutrino flavours have almost the same momentum distribution function, apart from small corrections,
and that the off-diagonal components of the neutrino density matrix are always very small,
all the flavour-changing NSI parameters are expected to affect \Neff\ in the same way.

\begin{figure}[t]
    \centering
    \includegraphics[width = 0.49\textwidth]{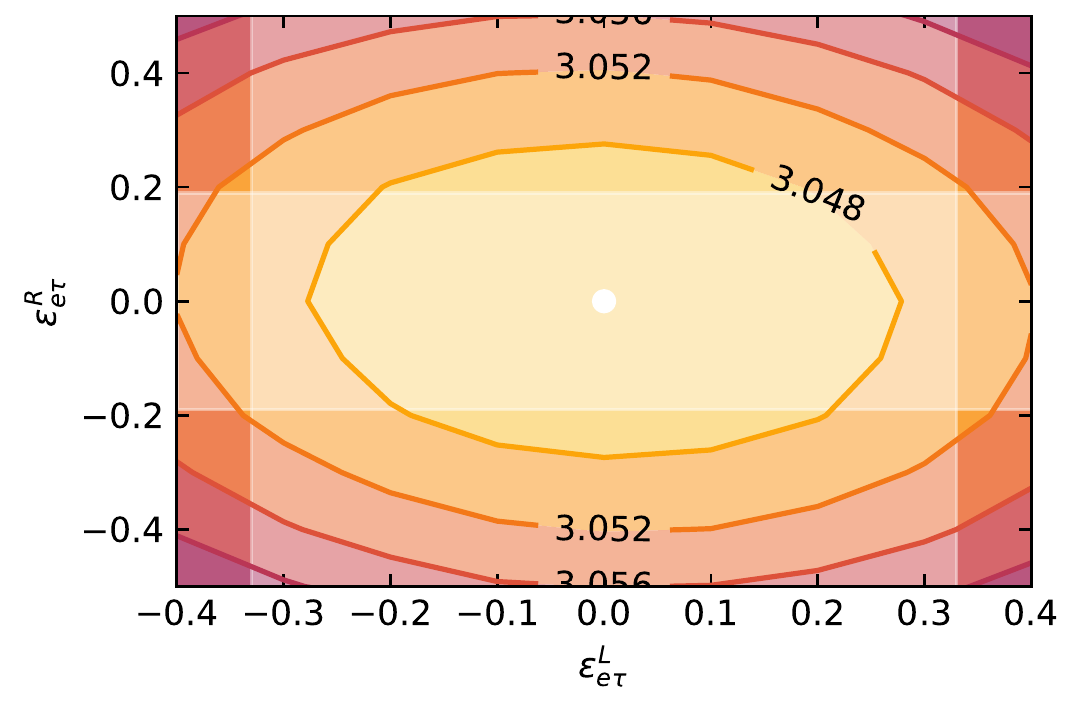}
    \includegraphics[width = 0.49\textwidth]{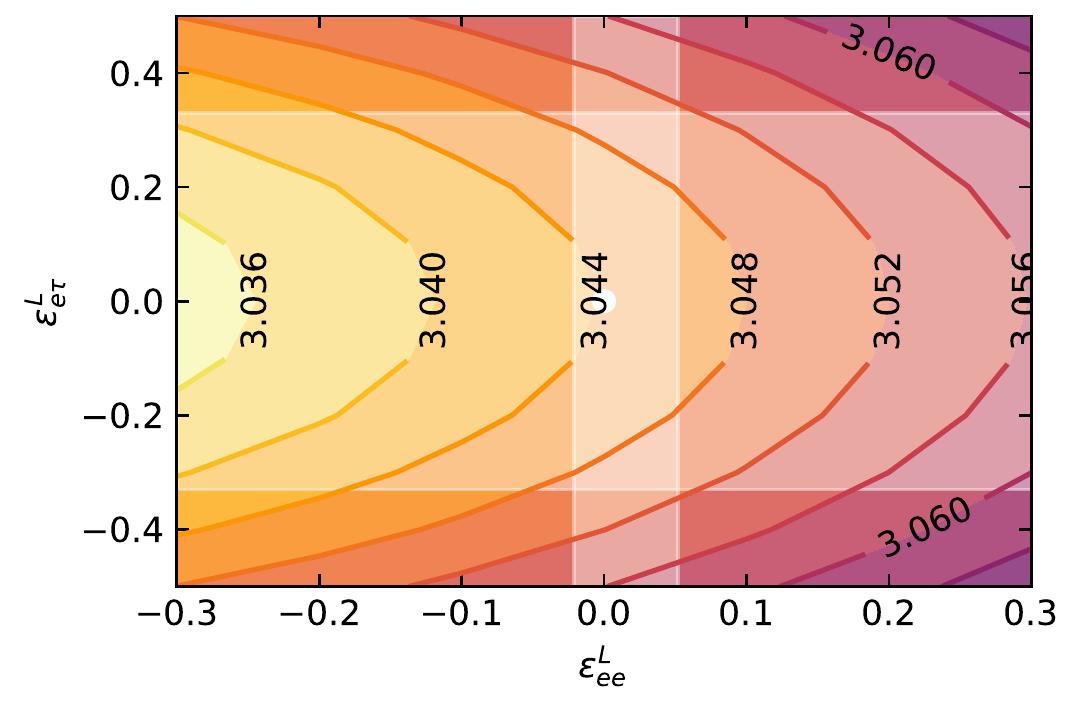}
    \includegraphics[width = 0.49\textwidth]{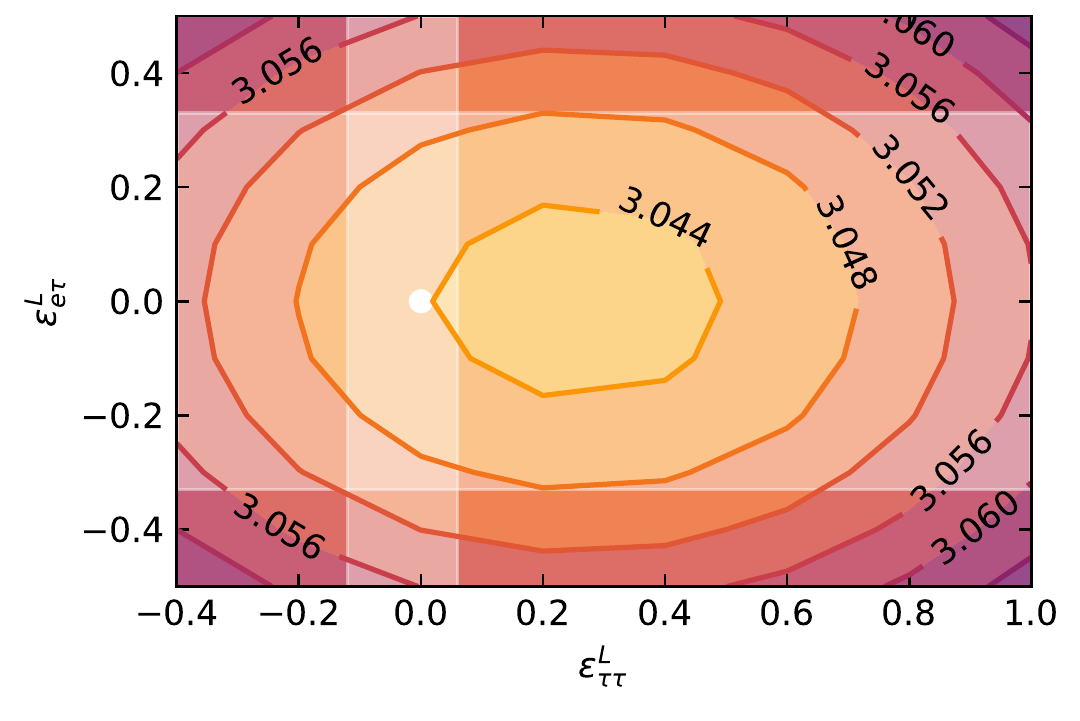}
        \includegraphics[width = 0.49\textwidth]{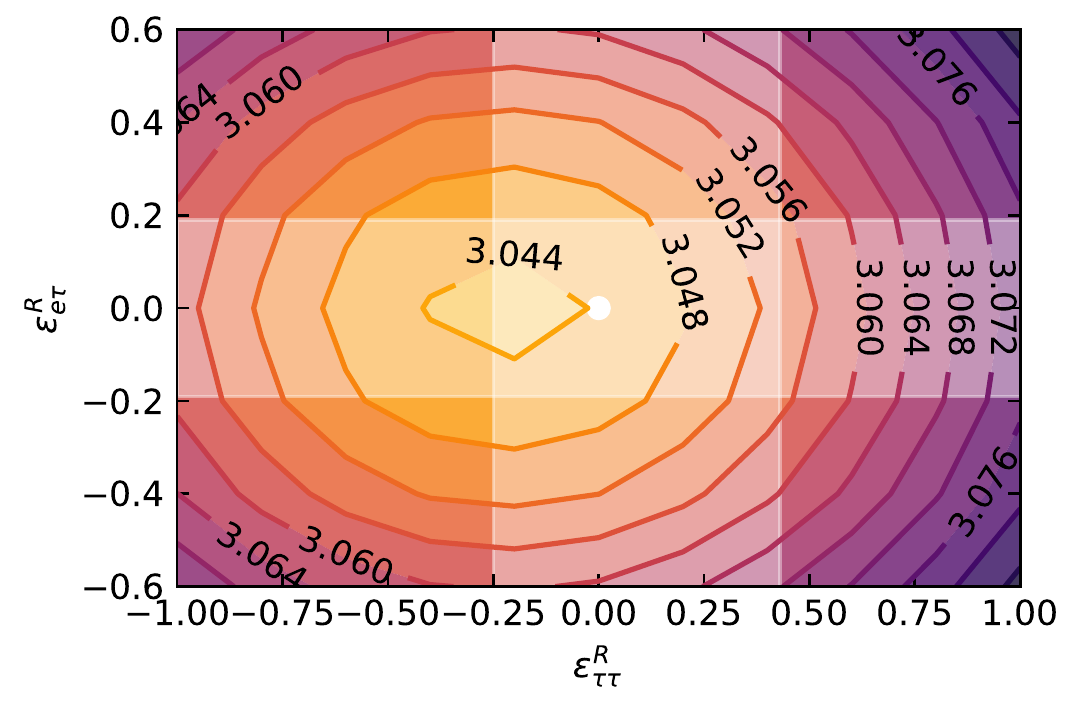}
    \caption{Same as Fig.\ \ref{fig:2dnu} for four possible combinations of two NSI parameters: $\varepsilon^L_{e\tau}$- $\varepsilon^R_{e\tau}$ (top left), $\varepsilon^L_{ee}$- $\varepsilon^L_{e \tau}$ (top right), $\varepsilon^L_{\tau\tau}$- $\varepsilon^L_{e\tau}$ (bottom left) and $\varepsilon^R_{\tau\tau}$- $\varepsilon^R_{e\tau}$ (bottom right). Again, white-shaded regions correspond to the 90\% CL experimental bounds obtained varying one parameter at a time, extracted from \citep{Farzan:2017xzy}.}
    \label{fig:2dfc}
\end{figure}

\begin{figure}[t]
    \centering
    \includegraphics[width = 0.49\textwidth]{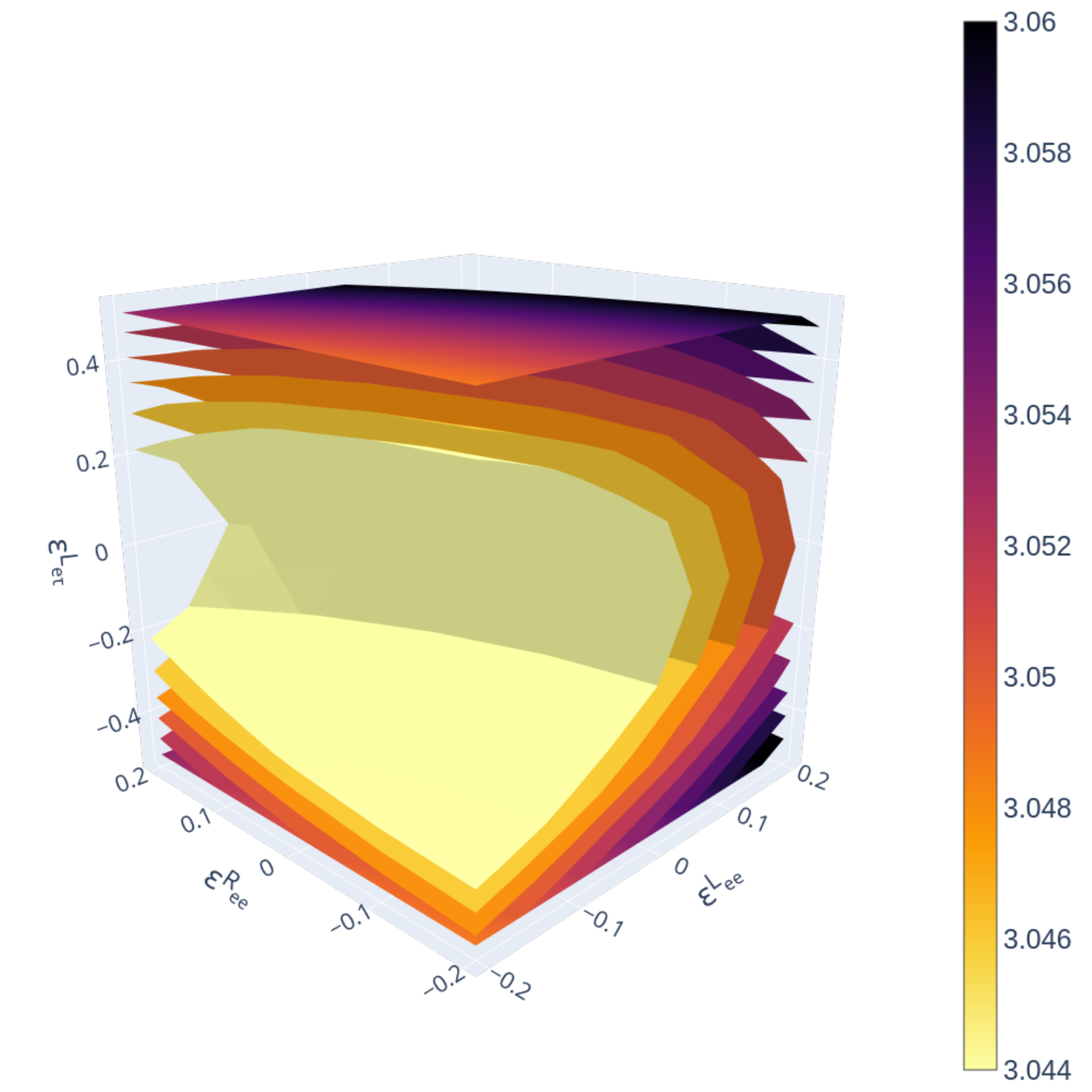}
    \includegraphics[width = 0.49\textwidth]{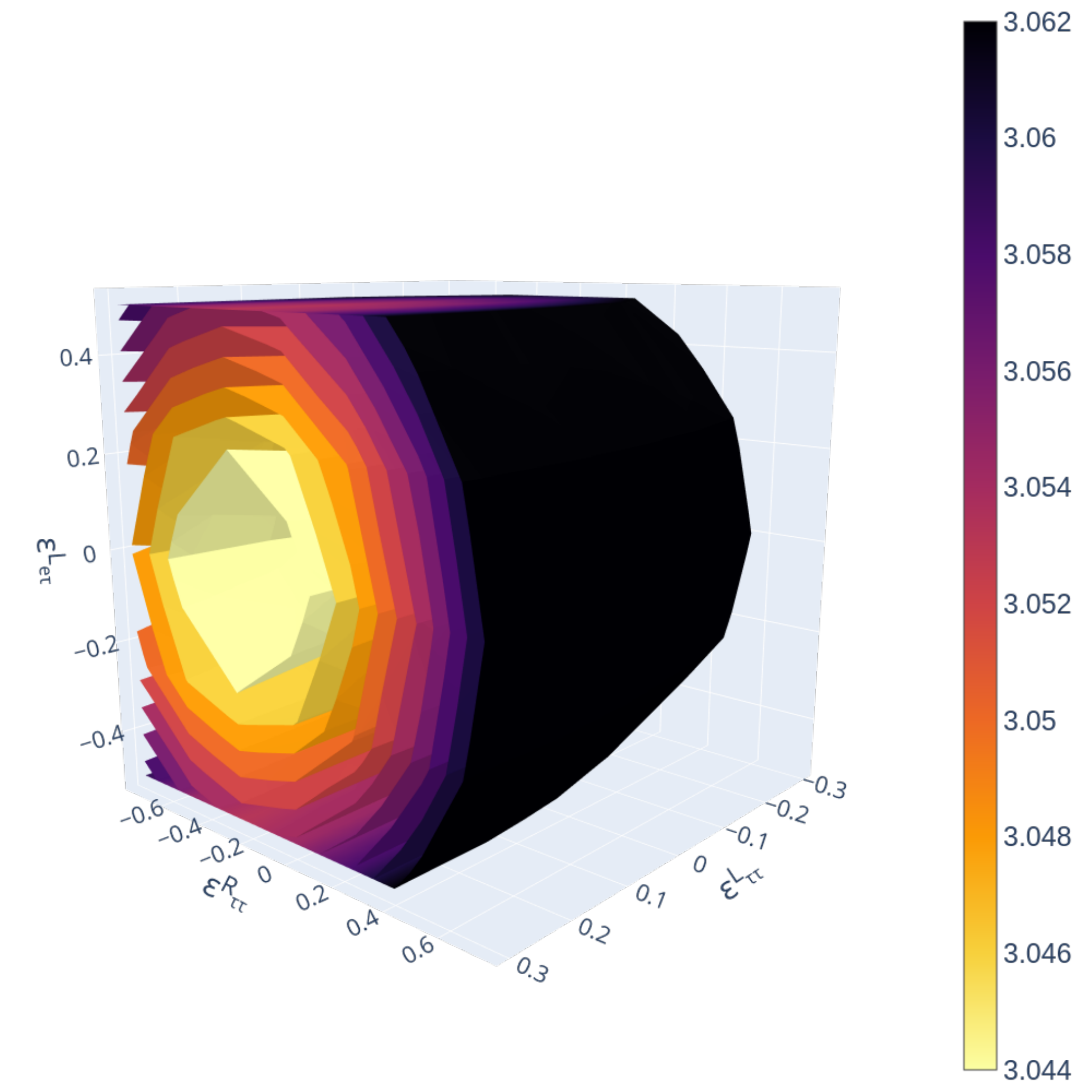}
    \caption{ Isosurfaces of $\Neff$ obtained from the simultaneous variation of three NSI couplings.}
    \label{fig:Neff3eps}
\end{figure}

The interplay between two flavour-changing NSI components or between non-universal and flavour-changing NSI terms can give rise to degeneracies as presented in Figure \ref{fig:2dfc}.
Moreover, when varying three NSI coefficients at the same time, we get ellipsoids, as shown in Figure~\ref{fig:Neff3eps} \footnote{An interactive version of Figure~\ref{fig:Neff3eps} in html format, that can be rotated, is available at \url{https://www.astroparticles.es/NSI_Neff/GL11_GR11_GL13.html} (left panel) or \url{https://www.astroparticles.es/NSI_Neff/GL33_GR33_GL13.html} (right panel).}.
The ellipses/ellipsoids can be understood based on the shift in the neutrino-electron couplings that NSI parameters introduce, as previously discussed.
These figures show how a precise determination of $\Neff$ could provide complementary information to improve the constraints on NSI, and demonstrate that an analysis varying more than one parameter at a time is feasible. 
From the figures we can notice that
considering several non-zero NSI parameters can lead to a value of $\Neff$ significantly larger than 3.044.
For instance, for $\varepsilon^L_{\tau \tau} = -0.60$, $\varepsilon^R_{\tau \tau} = -0.36$, $\varepsilon^L_{e\tau} = 0.132$, $\varepsilon^R_{e\tau} = -0.80$, $\varepsilon^L_{\mu \tau}$ = $\varepsilon^R_{\mu\tau}$ = 0.52, one obtains $\Neff = 3.10$.
Such a large deviation from the prediction in the absence of NSI could be tested by future determinations of $\Neff$ with a significance around 2--3$\sigma$.
The values of the parameters needed to reach such a significant deviation from $\Neff = 3.044$ are excluded by terrestrial experiments with a very high confidence level, although constraints are commonly derived varying one parameter at a time, therefore not taking into account the degeneracies between the various parameters.
In this case, cosmology provides access to a combination that is very unlikely to be proved on Earth (one would need, for instance, a beam of tau neutrinos in order to study $\nu_\tau$ scattering on electrons). Nonetheless, the expected constraints are not very competitive: they would rather be useful to provide an independent probe of the validity of terrestrial constraints than to provide strong bounds on the parameters themselves.

\section{Conclusions}
In this letter, we have addressed the impact of neutrino non-standard interactions with electrons on their decoupling from the primordial plasma. 
We have computed the evolution of neutrinos in the MeV epoch with non-zero NSI, using a modified version of the {\tt FortEPiaNO} code. Out of
the two main effects caused by NSI, namely the modifications to flavour oscillations in the dense QED medium and their contribution to scattering and annihilation processes between electrons and neutrinos, we have shown that the latter dominates. Nonetheless, a full treatment of neutrino oscillations with NSI was included in our analysis.

Precision calculations of the effective number of neutrinos, $N_{\rm eff}$, were performed, for the first time, for broad ranges of the NSI couplings with electrons $\varepsilon^{L,R}_{\alpha\beta}$, both for a single parameter and for multiparameter combinations.  
We found that, while non-universal NSI can either increase or decrease the energy transfer between neutrinos and electrons, leading to a value of $N_{\rm eff}$ that could be slightly smaller or larger than the one predicted by the Standard Model, flavour-changing NSI would only enhance \Neff.
For combinations of NSI parameters, there exist degeneracies leading to similar values of \Neff\, which can be well understood by the modifications that non-standard interactions introduce in the collision integrals. 

Our predictions for $N_{\rm eff}$ can be confronted with the forecasted sensitivity on this cosmological parameter from forthcoming 
CMB observations. Future bounds on NSI parameters from cosmological data are not expected to be, in general, competitive with present cosmological constraints from terrestrial experiments. However, a sensitivity of $\sigma(\Neff)=0.02-0.03$  \cite{Abazajian:2016yjj} could constrain non-universal NSI (e.g.\ $\varepsilon^R_{\tau\tau}$) to the same level as current laboratory experiments, or could bound some particular combinations of two (or more) NSI parameters leading to $\Neff\simeq 3.08$ or larger. In any case, cosmological observations provide an independent and complementary test on the existence of non-standard interactions of neutrinos with electrons.

To conclude, let us remark that slightly different and most promising results might be obtained in the context of different scenarios. In particular, for models imposing stronger conditions on the  NSI parameters as, for instance, neglecting the flavour dependence of the four-fermion operators giving rise to non-standard interactions, cosmology can provide stronger bounds on NSI~\cite{Du:2021idh}. 

\section*{Acknowledgments}
PFdS is supported by the Vetenskapsr{\aa}det (Swedish Research Council) through contract No.\ 638-2013-8993 and the Oskar Klein Centre for Cosmoparticle Physics.
SG acknowledges financial support from the European Union's Horizon 2020 research and innovation programme under the Marie Skłodowska-Curie grant agreement No 754496 (project FELLINI).
PMM, SP and MT acknowledge support from the Spanish grants FPA2017-85216-P (AEI/FEDER, UE) and PROMETEO/2018/165 (Generalitat Valenciana).
PMM is supported by the grant FPU18/04571.

\bibliography{bibliography}

\end{document}